# Near-resonant Raman amplification in the rotational quantum wavepackets of nitrogen molecular ions generated by strong field ionization

Zhaoxiang Liu[1,3], Jinping Yao[1,†], Jinming Chen[1,3,4], Bo Xu[1,3], Wei Chu[1], and Ya Cheng[1,2,5,*]

[1]*State Key Laboratory of High Field Laser Physics, Shanghai Institute of Optics and Fine Mechanics, Chinese Academy of Sciences, Shanghai 201800, China*

[2]*State Key Laboratory of Precision Spectroscopy, East China Normal University, Shanghai 200062, China*

[3]*University of Chinese Academy of Sciences, Beijing 100049, China*

[4]*School of Physical Science and Technology, ShanghaiTech University, Shanghai 200031, China*

[5]*Collaborative Innovation Center of Extreme Optics, Shanxi University, Taiyuan, Shanxi 030006, China*

[†]*jinpingmrg@163.com*

[*]*ya.cheng@siom.ac.cn*

# Abstract

Generation of laser-like narrow bandwidth emissions from nitrogen molecular ions ($N_2^+$) generated in intense near- and mid-infrared femtosecond laser fields has aroused much interest because of the mysterious physics underlying such a phenomenon as well as the potential application of such an effect in atmospheric spectroscopic sensing. Here, we perform a pump-probe measurement on the nonlinear interaction of rotational quantum wavepackets of $N_2^+$ generated in mid-infrared (e.g., at a wavelength centered at 1580 nm) femtosecond laser fields with an ultrashort probe pulse whose broad spectrum overlaps both P- and R-branch rotational transition lines between the electronic states $N_2^+(B^2\Sigma_u^+, v'=0)$ and $N_2^+(X^2\Sigma_g^+, v=0)$. The results show that in the near-resonant conditions, stimulated Raman amplification can efficiently occur which converts the broad bandwidth ultrashort probe pulse into the narrow bandwidth laser-like beam. Our finding provides an insight into the physical mechanism of strong field induced lasing actions in atmospheric environment.

Non-perturbative nonlinear optics underpinned by strong field tunnel ionization has opened the door to high-order harmonic generation and attosecond science [1-3]. Recently, it is observed that the non-perturbative interaction of intense laser fields with molecules can enable to generate laser-like narrow-bandwidth coherent emissions instantly after the photoionization [4-19]. The wavelength of each narrow-bandwidth emission is in accordance with one of transitions between the vibrational energy levels of excited $N_2^+(B^2\Sigma_u^+)$ state and that of ground $N_2^+(X^2\Sigma_g^+)$ state [4]. To understand the intriguing phenomenon, pump-probe measurements have been earlier carried out by setting the pump wavelength at ~800 nm and the probe wavelength around 391 nm, corresponding to the transition between $N_2^+(B^2\Sigma_u^+, v'=0)$ and $N_2^+(X^2\Sigma_g^+, v=0)$ [7,8,10-14,16,17]. The time window of gain, which covers multiple revival periods of rotational wavepackets in $N_2^+$ ions, provides clear evidence on the population inversion between excited and ground states. At present, several mechanisms have been proposed for understanding the origin of the population inversion generated in the 800 nm laser field, including field-induced multiple recollisions [12,18], depletion of ground state via one-photon absorption [14,15], and transient inversion between rotational wavepackets of upper and lower states [19], etc. Nevertheless, a reliable mechanism is yet to be identified.

Interestingly, generation of the laser-like narrow-bandwidth coherent emissions appears sensitive to the pump wavelength [4,14,18], whereas a pump-probe measurement has never been performed at pump wavelengths other than 800 nm. To shed more light on this characteristic, we perform pump-probe measurements on the generation of narrow bandwidth emission at ~391 nm wavelength in $N_2^+$ ions with mid-infrared pump pulses whose wavelength is centered at 1580 nm wavelength. Specifically, the probe pulse is generated to have an initial spectrum which covers both P and R branches of the rotational transition spectrum between $N_2^+(B^2\Sigma_u^+, v'=0)$ and $N_2^+(X^2\Sigma_g^+, v=0)$.

The dependence of the spectrum of probe pulses on the pump-probe time delay reveals that the nonlinear interaction of light and coherent rotational quantum wavepackets of molecular ions in the near-resonant conditions can play a crucial role in converting the broad bandwidth probe pulse into a narrow bandwidth laser-like coherent emission.

The experiment was carried out using an optical parametric amplifier (OPA, HE-TOPAS, Light conversion Ltd.), which was pumped by a commercial Ti:sapphire laser system (Legend Elite-Duo, Coherent, Inc.). The OPA enables us to generate 1580 nm, 0.92 mJ, ~60 fs laser pulses at a repetition rate of 1 kHz, which serves as the pump to ionize nitrogen molecules and meanwhile, creates the coherent rotational wavepackets of molecular ions. The beam diameter ($1/e^2$ width) is measured to be ~7.6 mm. The probe pulse with a pulse energy of ~20 nJ was generated through frequency doubling of the residual 800 nm femtosecond laser pulse from OPA, whose spectrum is shown in Fig. 2 (gray dotted curve). Thus, the probe pulse can resonantly interact with $N_2^+$ ions via the transition between the ground $N_2^+(X^2\Sigma_g^+)$ and excited $N_2^+(B^2\Sigma_u^+)$ states. A polarizer was inserted in the optical path of the probe beam to ensure a highly linear polarization. The polarization of pump pulse is set to be parallel to that of the probe. After being combined by a dichroic mirror, the pump and probe beams were collinearly focused by an $f$=15 cm lens into a gas chamber filled with nitrogen gas at a pressure of 20 mbar. The time delay between the two was controlled by a motorized translation stage. The zero delay is defined as the temporal overlap of two beams, which was determined by maximizing the sum frequency generation in a BBO crystal. The spatial overlap of the two beams was determined by looking for the strongest absorption in the probe spectrum. The pump and probe beams exiting from the gas chamber were first separated by a dielectric mirror with a high reflectivity in the 350~400 nm wavelength range, and then the residual pump was further eliminated using a piece of blue glass. At last, the probe pulses was focused by a lens into an imaging spectrometer (Shamrock 500i, Andor) with a 2400 lines/mm grating for spectral analysis.

First, we measured the spectrum of probe pulse as a function of pump-probe time delay. As shown in Fig. 1, when the pump lags behind the probe (i.e., negative delay), the spectrum of probe pulse is hardly affected by the pump beam. In contrast, when the pump is ahead of the probe (i.e., positive delay), a series of fine absorption dips appear in the spectrum of the probe. The prominent absorption band close to 391 nm corresponds to the P branch of rotational transition $N_2^+(X^2\Sigma_g^+, v=0, J) \rightarrow N_2^+(B^2\Sigma_u^+, v'=0, J')$ (i.e., $J'-J=-1$), whereas a series of discrete absorption dips on the blue side of 391 nm are in accordance with the R branch of rotational transition (i.e., $J'-J=1$). Here, J and J′ denote the rotational quantum numbers of $N_2^+(X^2\Sigma_g^+, v=0)$ state and $N_2^+(B^2\Sigma_u^+, v'=0)$ state, respectively. For clarity, the rotational quantum numbers of $X^2\Sigma_g^+(v=0)$ state related to R-branch transitions are also labeled on the spectrum in Fig. 1. The J numbers are calculated using the two rotational constants of $X^2\Sigma_g^+(v=0)$ (i.e., $B=1.9223$ cm$^{-1}$) and $B^2\Sigma_u^+(v'=0)$ (i.e., $B'=2.0738$ cm$^{-1}$) states [20].

From the measured spectrum, we determine that some rotational quantum states as high as $J>30$ can be generated in $N_2^+(X^2\Sigma_g^+, v=0)$ state with our pump laser field. The measured spectrum also shows that the absorption from even-order rotational energy levels is stronger than that from the odd-order levels, which is a result of nuclear spin statistics [21]. Additionally, two strong absorption lines have been observed at the wavelengths of 388.4 nm and 385.8 nm, which can be attributed to transitions of $N_2^+(X^2\Sigma_g^+, v=1) \rightarrow N_2^+(B^2\Sigma_u^+, v'=1)$ and $N_2^+(X^2\Sigma_g^+, v=2) \rightarrow N_2^+(B^2\Sigma_u^+, v'=2)$, respectively. These results show that strong field ionization of molecules is capable of creating a broad distribution of rotational states in ground state of $N_2^+$ ions. Interestingly, it was observed that R-branch absorption spectra show pronounced fast oscillations with the increase of pump-probe delay. The oscillation frequency increases with the rotational quantum number J, and thus some arc-shaped structures (indicated by the white dotted line in Fig. 1 for the guide of eyes) form in the time-resolved absorption spectrum. Furthermore, absorptions from different rotational quantum states simultaneously reach maximum or minimum at time delays near each half and full revival period of $N_2^+(X^2\Sigma_g^+)$ state.

We now take a close look at the spectral structures near the first full revival period. For comparison, we chose three specific time delays $t_0 = -0.3363$ ps, $t_1 = 8.6000$ ps and $t_2 = 8.7670$ ps as indicated by the arrows in Fig. 1. The absorption efficiencies of all the R-branch rotational transitions reach either minimum (at time delay $t_1$) or maximum (at time delay $t_2$) regardless of the quantum number J. The spectrum of probe laser at $t_0$ is almost unchanged as compared with its original spectrum measured before the gas chamber, and thus can be used as a reference. Figure 2 shows the typical spectra measured at the three time delays. To facilitate our analysis, the spectra in Fig. 2 are divided into three regions, i.e., the R branch (region I) and P branch of rotational transitions (region II), and the region on the red side of the bandhead of P-branch transitions (region III). It can be observed that at $t_2$, a gross loss appears in the spectral range from 386 nm to ~391 nm in region I as compared to the reference spectrum (gray dotted line), while the signal of probe pulse has been significantly enhanced at some specific wavelengths of regions II and III. A completely opposite situation has been observed in the spectrum recorded at $t_1$ as shown by the red dash dot curve in Fig. 2. In this case, the probe pulse remains almost unchanged in the spectral range from 386 nm to 390 nm except for those absorption lines corresponding to the resonant rotational transitions, while a strong absorption in the probe pulse can be observed in regions II and III.

To reveal the origin of observed fast oscillations in Fig. 1, we perform Fourier transforms of the R-branch absorption spectra. As illustrated in Fig. 3(a), the Fourier spectrum for each R-branch rotational transition mainly includes two frequency components, i.e., $(4J-2)Bc$ and $(4J+6)Bc$, as indicated by the dashed line and the dotted line in Fig. 3(a). The two frequencies respectively correspond to the beat frequency between J and J−2 states and that between J and J+2 states of $N_2^+(X^2\Sigma_g^+, v=0)$ [11,17]. The linear dependence of the quantum beats on the rotational quantum number J results in the arc-shaped structures presented in Fig. 1. This beat frequency originates from the modulation of populations in the individual rotational energy levels of $N_2^+(X^2\Sigma_g^+, v=0)$ state caused by the resonant interaction of probe pulses with $N_2^+$ ions. In brief, the coherent rotational wavepackets of $N_2^+$ ions in the ground state are initially created by the nonadiabatic interaction between the intense pump laser with nitrogen molecules along with the photoionization. Afterwards, each rotational quantum state will evolve freely with a time-dependent phase $E_J t + \phi_J$, where $E_J$ is rotational energy of J state and $\phi_J$ the initial quantum phase in the field-free evolution. Then, the delayed probe pulse will modulate the population distributions in different rotational states via resonant rotational Raman scattering. This process naturally occurs along with the one-photon absorption because the spectrum of the probe pulse covers both P- and R-branch transitions. The final population in each rotational J state oscillates with the increasing pump-probe delay at two main oscillation frequencies $(4J-2)Bc$ and $(4J+6)Bc$, as discussed in our previous work [17]. The population modulation will give rise to the time-dependent R-branch absorption spectra as presented in Fig. 1.

We now look at the spectral characteristics in region II which can be attributed to P

branch of the rotational transitions. The theoretically calculated absorption wavelengths of P(J) photons are indicated by the black-dashed vertical lines and red-dotted vertical lines in Fig. 3(b). Clearly, the dips in the spectrum are in good agreement with P-branch absorption lines for the relatively low J states (i.e., vertical black dashed line). Similar to the R-branch absorption spectra, the nuclear spin statistics results in the absorption from even-order J state is stronger than that from the odd-order J state for P-branch rotational transitions. At the time delay $t_2$, the signal in region II is enhanced in comparison with the reference spectrum of the probe pulse (i.e., the gray dotted line), whereas the signal in region I is mostly suppressed due to the loss. The fact indicates an energy transfer between the two regions. From the definition of P- and R-branch rotational transitions, the absorption of one R(J) photon at the frequency of $\omega_R(J)$ and the subsequent emission of one P(J+2) photon at the frequency of $\omega_P(J+2)$ synergetically give rise to a Stokes process of Raman scattering in $N_2^+$ ions, as illustrated in Fig. 4(a). Likewise, the absorption of one P(J) photon and the sequent emission of one R(J−2) photon constitute an anti-Stokes process. As a result, the energy transfer between P- and R-branch signals should be a result of rotational Raman scattering caused by the interaction between the probe pulse and rotational wavepackets of molecular ions.

Naturally, one may argue that the probe used in our experiment is too weak to generate such strong stimulated Raman scattering (SRS). To understand this, it should be noticed that the weak probe pulse covering both R- and P-branch transitions, the Stokes and anti-Stokes photons involved in the Raman process are all in resonance with the rotational transitions between $N_2^+(X^2\Sigma_g^+, v=0)$ and $N_2^+(B^2\Sigma_u^+, v'=0)$. The resonant interaction efficiently promotes the gain and lowers the threshold of SRS [22]. On the other hand, the intense mid-infrared laser field has initiated impulsive Raman excitation prior to the arrival of the probe, resulting in the generation of coherent rotational wavepackets composed of $J+2n$ ($n = 0, \pm 1, \pm 2 \cdots$) rotational energy levels of $N_2^+(X^2\Sigma_g^+, v=0)$ state [23]. Subsequent resonant interaction of rotational wavepackets with a weak probe field naturally leads to strong SRS around the resonance wavelengths.

Surprisingly, a strong emission appears at wavelengths longer than 391.42 nm and extends to nearly 392 nm at time delay $t_2$. This spectral region is beyond the bandhead of P-branch transitions (i.e., 391.42 nm), and thus it cannot be reached by any resonant transitions in the R and P branches. For clarity, we zoom into the spectral range from 390.9 nm to 392.5 nm in Fig. 3(c). Clearly, the spectrum shows a multiple-plateau structure. The origin of the off-resonant emission peaks in region III are very intriguing. We notice in Fig. 2 that the off-resonant emissions in region III become most pronounced when the absorption (i.e., the loss) in region I is the strongest. The physics underlying such an observation is schematically depicted in Fig. 4(a) and (b). From the point of view of near-resonant rotational Raman scattering, the absorption of one photon with the frequency $\omega_R(J) < \omega < \omega_R(J+1)$ will lead to the production of one Stokes photon with the frequency $\omega - \Omega_{J+2}^J$ or $\omega - \Omega_{J+3}^{J+1}$. Here, $\Omega_{J+2}^J$ and $\Omega_{J+3}^{J+1}$ are the rotational frequency difference between $J+2$ and $J$ states and that between $J+3$ and $J+1$, respectively, which follow $\Omega_{J+2}^J = (4J+6)Bc$ and $\Omega_{J+3}^{J+1} = (4J+10)Bc$. The limit of up-shifted wavelength, as indicated by the pink crosses in Fig. 4(c), is determined by the first near-resonant Raman process in Fig. 4(a). Namely, the photon at the frequency of $\omega_R(J)$ inelastically scattered from $J+1$ state of $N_2^+(X^2\Sigma_g^+, v=0)$ will give rise to the SRS signal at the frequency of $\omega_R(J) - \Omega_{J+3}^{J+1}$. In contrast, the limit of the down-shifted wavelength, as indicated by the black squares in Fig. 4(c), is determined by the process in Fig. 4(b). In this case, all the R(J) photons inelastically scattered from $J-1$ energy level of $N_2^+(X^2\Sigma_g^+, v=0)$ will give rise to the SRS signal at the frequency of $\omega_R(J) - \Omega_{J+1}^{J-1}$. As a result, the wavelengths of Stokes signals generated by the rotational Raman scattering will distribute in the region between the pink crosses and black squares in Fig. 4(c). We stress that the generation of near-resonance SRS signals can be very efficient as the involved wavelengths are out of the absorption range of $N_2^+(X^2\Sigma_g^+, v=0)$. This off-resonant condition ensures that the gain in region III will not be suppressed by any single photon absorption.

Quantitatively, as shown in Fig. 4(a) and (c), the first near-resonant stimulated Raman amplification process can generate Stokes photons with the longest wavelength of 391.54 nm as indicated by vertical dotted line, which agrees very well with the observed first plateau in Fig. 3(c). For the R(J) photons inelastically scattered from the higher rotational energy levels, the Stokes photons will be produced at longer wavelengths, because the rotational energy difference increases with the rotational quantum numbers. For the R(J) photons scattered from J+2, J+3, J+4, J+5 rotational states, the calculated boundary wavelengths of the corresponding near-resonant Raman scattering processes are indicated by the dotted lines in Fig. 3(c). Remarkably, all the plateaus in the measured spectrum and the calculation results show excellent agreement. Furthermore, with the increase of emission wavelength, these near-resonant processes have lower efficiencies due to the increased energy detuning. The quantitative evidence unambiguously confirms the occurrence of highly efficient near-resonant SRS in $N_2^+$ ions.

To sum up, we have performed a pump-probe measurement on the nonlinear interaction of strong-field-excited rotational quantum wavepackets of $N_2^+$ with a weak ultrashort probe pulse. The unambiguous evidence on the near-resonant coherent Raman scattering provides an insight on the mechanism of laser-like emissions generated in air with mid-infrared pump pulses. The efficient nonlinear process in the near ultraviolet region not only provides a means to practice nonlinear optics at short wavelengths, but also opens the prospect of nonlinear spectroscopy in atmospheric environment.

This work is supported by the National Basic Research Program of China (Grant No. 2014CB921303), National Natural Science Foundation of China (Grant Nos. 61575211, 11674340, 61405220 and 11404357), the Strategic Priority Research Program of Chinese Academy of Sciences (Grant No. XDB16000000), Key Research Program of Frontier Sciences, Chinese Academy of Sciences (Grant No. QYZDJ-SSW-SLH010), and Shanghai Rising-Star Program (Grant No. 17QA1404600).

**Captions of figures:**

Fig. 1 (Color online) Spectra of the probe pulse exiting from the gas chamber as a function of the time delay between pump and probe pulses.

Fig. 2 (Color online) The spectra of probe pulse measured at time delays $t_0$ (gray dot line), $t_1$ (red dash dot line) and $t_2$ (blue solid line).

Fig. 3 (a) Fourier transforms of the R-branch absorption spectra in Fig. 1(a), and the zoom-in spectra of Fig. 2 in the spectral ranges (b) and (c) fitted with theoretical calculations.

Fig. 4 (Color online) (a) Schematic diagram of resonant and the first near-resonant Raman scattering in nitrogen molecular ions. (b) Schematic diagram of resonant and the second near-resonant Raman scattering in nitrogen molecular ions. (c) Calculated Stokes wavelengths for the near-resonant Raman processes in (a) and (b).

Fig. 1

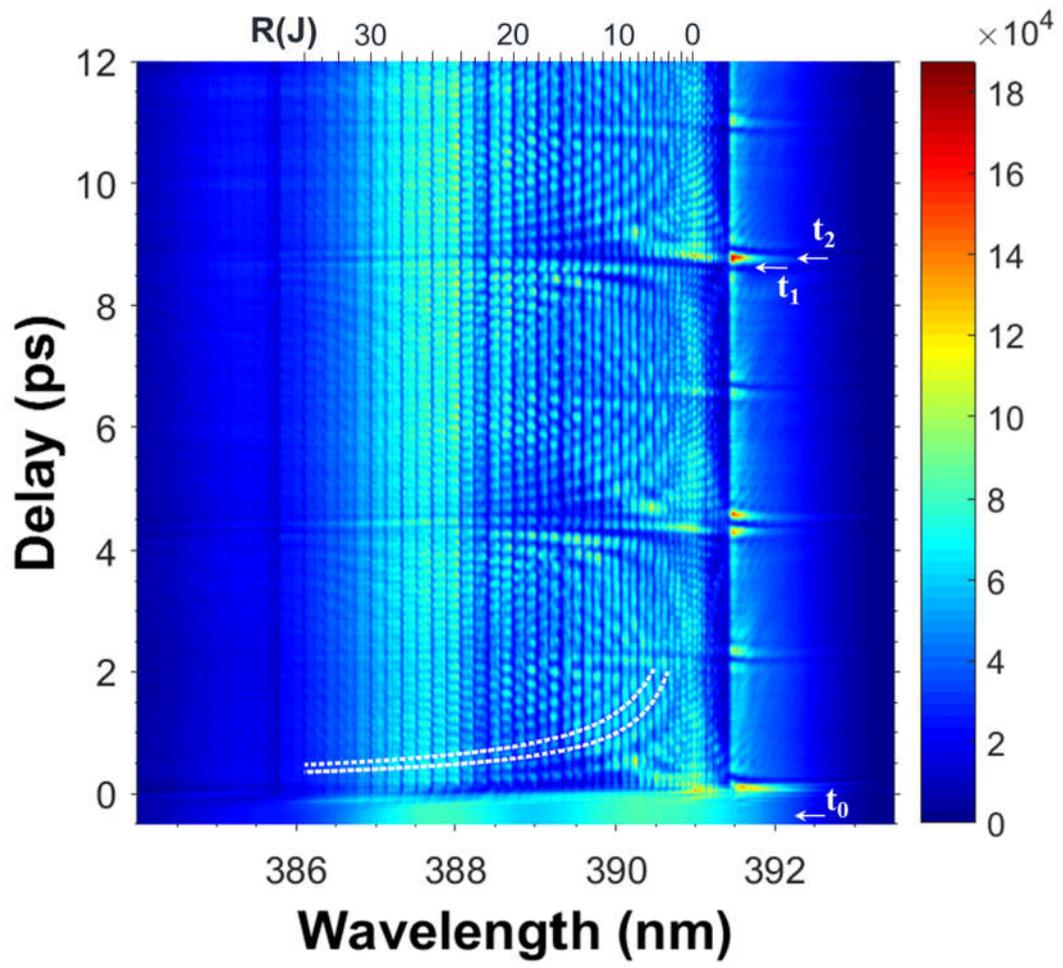

Fig. 2

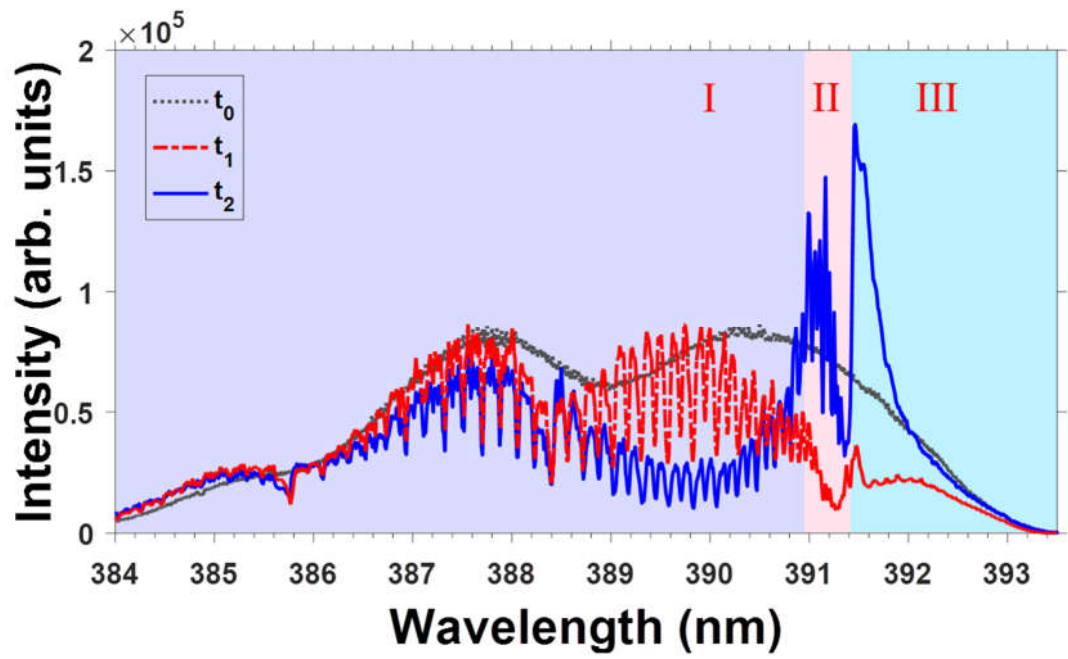

Fig. 3

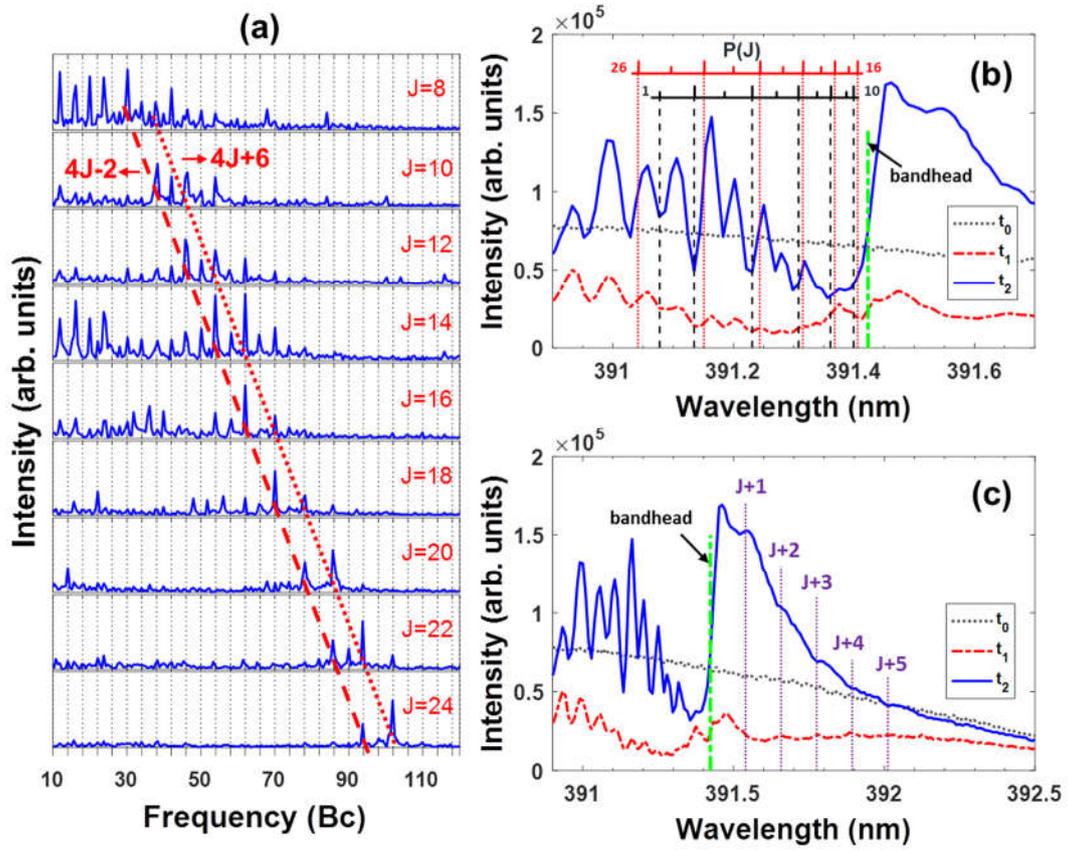

Fig. 4

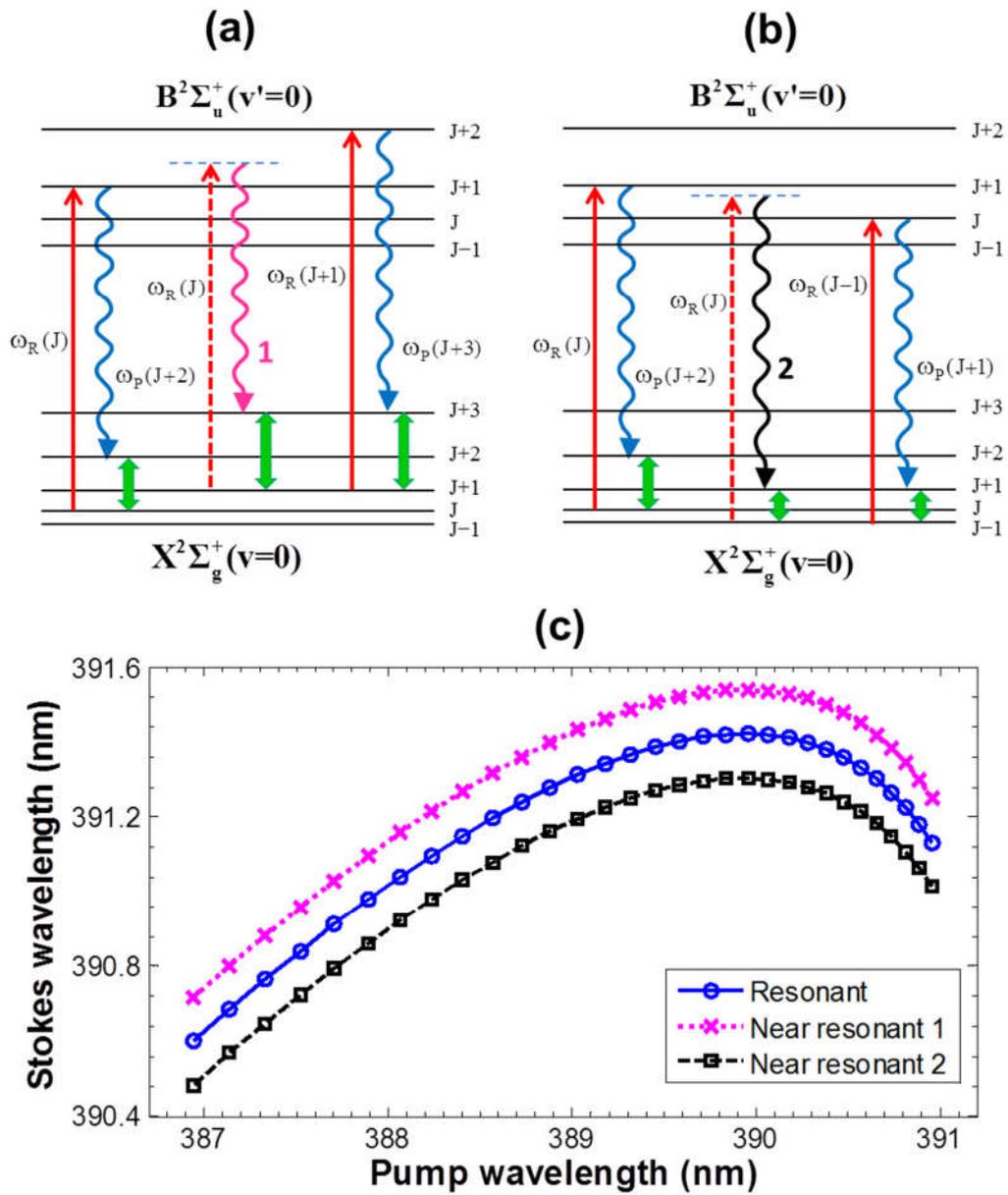